\documentclass[aps,prl,twocolumn,showpacs]{revtex4b5}

\usepackage{amsmath}
\usepackage{graphicx}

\allowdisplaybreaks

\begin{document}

\title{Uniting Bose-Einstein condensates in optical resonators}
\author{D.~Jaksch$^1$, S.A.~Gardiner$^{1,2}$ K.~Schulze$^1$, J.I.~Cirac$^1$,
and P.~Zoller$^1$} \affiliation{${}^1$ Institut f\"{u}r
Theoretische Physik, Universit\"at Innsbruck, A--6020 Innsbruck,
Austria.} \affiliation{${}^2$ Institut f\"{u}r Physik,
Universit\"{a}t Potsdam, D--14469 Potsdam, Germany}

\begin{abstract}
The relative phase of two initially independent Bose-Einstein
condensates can be laser cooled to unite the two condensates by
putting them into a ring cavity and coupling them with
an internal Josephson junction. First, we show that this phase
cooling process already appears within a semiclassical model. We
calculate the stationary states, find regions of bistable behavior
and suggest a Ramsey-type experiment to measure the build up of
phase coherence between the condensates. We also study quantum
effects and imperfections of the system.
\end{abstract}
\pacs{03.75.Fi, 74.50.+r, 42.50.-p}

\maketitle

During recent years Bose-Einstein condensates (BEC) of Alkali
atoms and Hydrogen have been produced and studied extensively in
the laboratory \cite{experiments}. In view of potential
applications, such as the generation of bright beams of coherent
matter waves (atom laser), a central goal has been the formation
of condensates with a number of atoms as large as possible. It is
thus of particular interest to study a scenario where this goal is
achieved by uniting two (or more) independently grown condensates
to form one large single condensate. Physically speaking, two
independently formed condensates are characterized by a random
relative phase of their macroscopic wave functions. A ``fusion''
of two condensates thus amounts to locking the relative phase in a
dissipative process. Below we will study such a mechanism in the
context of optical Cavity QED \cite{QCED}. In our schemes two
condensates in different internal atomic states are coupled by
lasers in a Raman configuration (internal Josephson effect
\cite{JJ}), and in addition to a lossy optical cavity, which
provides an effective zero temperature reservoir.
As a result, we obtain a damping mechanism for the relative 
condensate phase. A physical picture behind this cooling mechanism 
can be given by establishing a formal analogy of the dynamical 
equations describing this ``laser cooling'' of the relative phase, 
and the recently discussed cavity assisted laser cooling of the 
motional degrees of freedom of atoms or molecules \cite{singleatomcool}.

We consider two equally large, independently produced BECs, where
the atoms of the individual condensates are the same species but
in two {\em different\/} hyperfine states $|1\rangle$ and
$|2\rangle$. The elements of the one-body density matrix are given
by $\rho_{kl}({\mathbf x},{\mathbf x}') = \langle
\psi_{l}^{\dagger}({\mathbf x}') \psi_{k}({\mathbf x})\rangle$
\cite{DensMat}. Here, $\psi_{k}({\bf x})$ is a bosonic field
operator which annihilates a particle at position ${\mathbf x}$
and in hyperfine state $|k\rangle$, and $k,l\in\{1,2\}$. The
one-body density matrix of these two independent condensates has
vanishing off diagonal elements $\rho_{12}({\mathbf x}, {\mathbf
x}')=0$, and two dominant eigenvalues of approximately $N/2$,
where $N$ is the total particle number. The diagonal terms can
then be written as $\rho_{kk}({\mathbf x}, {\mathbf x}')\approx N
\varphi^*_k({\bf x}') \varphi_k({\bf x})/2$, with $\varphi_k ({\bf
x})$ the wave function of the condensate in state $|k\rangle$.
There is thus no phase-coherence between the two different
condensates and any attempt to measure a relative condensate phase
$\Phi$ will produce a random result \cite{JJ}. Here we study the
possibility of building up and locking the phase $\Phi$ by (i) an
internal atomic Josephson junction (JJ) \cite{JJ,JILARams} transferring
atoms between the two initial condensates, and (ii) coupling the
atoms to a dissipative ring cavity \cite{Sols}. The final result is that for
experimentally relevant parameters the phase $\Phi$ develops
towards a stationary value on a time scale of a few trap
oscillation periods. This is reflected by one dominant eigenvalue
of approximately $N$ of $\rho_{kl} ({\mathbf x},{\mathbf x}')$.
The corresponding condensate mode may be an electronic
superposition state. In this sense we have thus joined two
condensates together to form one larger condensate.

\begin{figure}
\begin{center}
\includegraphics[width=7.7cm]{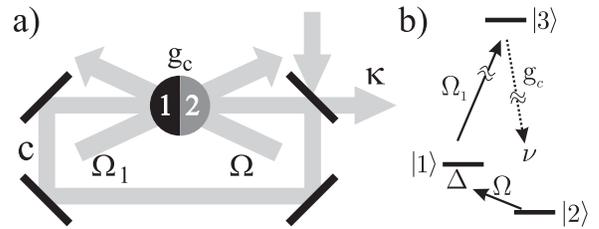}
\end{center}
\caption{a) Experimental setup: Two initially independent
condensates $1$ and $2$ are trapped in a ring cavity and coupled
to the cavity mode $c$ as well as to the two lasers $\Omega$ and
$\Omega_1$. b) Level structure: The Raman lasers $\Omega$ transfer
atoms from $|1\rangle$ to $|2\rangle$. A laser $\Omega_1$ drives
the transition between particles in $|1\rangle$ and an auxiliary
level $|3\rangle$. The cavity couples the levels $|3\rangle$ and
$|2\rangle$ with coupling strength $g_c$.} \label{fig1}
\end{figure}

The internal JJ \cite{JJ} is realized by
coupling the two hyperfine states $\{|1\rangle,|2\rangle\}$ by a
Raman transition with detuning $\Delta$ and effective Rabi
frequency $\Omega$ (cf.~Fig.~\ref{fig1}). We assume $\Omega$ to be
real, positive, and ${\mathbf x}$ independent, i.e.~there is
negligible momentum transfer to the condensates due to the Raman
lasers. In second quantized form the Hamiltonian $H_{\rm
BEC}=H_1+H_2+H_{L}$ \cite{JJ} is given by (with $\hbar=1$)
\begin{subequations}
\begin{align}
H_{k}=& \int \! d{\mathbf x} \; \psi^\dagger_{k} \left(
-\frac{\nabla^{2}}{2m}
+V+\sum_{l}\frac{u_{kl}}{2} \psi^\dagger_{l} \psi_{l} \right) \psi_{k},\\
H_{L}=& \int d{\mathbf x} \left[\Delta\psi^\dagger_1 \psi_1
+\left(\frac{\Omega}{2} \psi_1^\dagger \psi_2 + {\rm h.c.} \right)
\right],
\end{align} \label{H}
\end{subequations}
where $\psi_{1,2} \equiv \psi_{1,2}({\mathbf x})$, $V \equiv
V({\bf x})$ the trapping potential; we have suppressed ${\bf x}$
for notational convenience. We assume the trapping potential to be
harmonic with frequency $\omega$ and the mass of the atoms to be
$m$. Two-particle interactions are characterized by $u_{kk}=4 \pi
a_{k} /m$ where $a_{k}$ is the $s$-wave scattering length of atoms
in condensate $k$ and by $u_{12}=u_{21}=4 \pi a_{12} /m$ with
$a_{12}$ the interspecies $s$-wave scattering length. We restrict
ourselves to the case where $a_1 a_2 > a_{12}^2$, i.e.~to two
stable strongly overlapping condensates \cite{JJ}.

The ring cavity mode couples the state $|2\rangle$ to an auxiliary
excited atomic state $|3\rangle$, with detuning $\delta_c$ and
coupling strength $g_c({\bf x})$, which decays at a rate $\kappa$. We
also add a classical laser field driving the $|1\rangle$ to
$|3\rangle$ transition with Rabi frequency $\Omega_1({\bf x})$ and
detuning $\delta_1$ (cf.\ Fig.\ \ref{fig1}). Adiabatically
eliminating the internal state $|3\rangle$ (which requires $\delta_1 \gg
\Omega_1({\mathbf x})$ and $\delta_c \gg g_c({\mathbf x}) |C|$,
with $|C|$ the square root of the number of photons in the cavity) we
obtain the following master equation with $H=H_{\rm BEC}+H_C$:
\begin{equation}
\dot{\rho} = - i[H,\rho]
 +\frac{\kappa}{2}\left(2 c\rho c^{\dagger}
-c^{\dagger}c\rho - \rho c^{\dagger}c \right), \label{master}
\end{equation}
where $c$ is the annihilation operator of a photon in the cavity
mode, $\nu=\delta_c-\delta_1-\Delta$ is the effective cavity
detuning, and
\begin{equation}
H_{C} = \int \! d{\mathbf x} \left[g({\bf x}) c \psi_1^\dagger
\psi_2 + {\rm h.c.}\right] +\nu c^{\dagger}c.
\end{equation}
The effective coupling of the condensates to the cavity is given
by $g({\bf x})=\Omega_1({\bf x}) g_c({\bf x})/2 \delta_1$. In a
ring cavity the spatial dependences of $\Omega_1({\bf x})$ and
$g_c({\bf x})$ can be made to almost cancel each other. There is
thus negligible momentum transfer to the atoms due to the coupling
to the cavity, and we may set $g({\bf x})=g$, where $g$ is
${\mathbf x}$ independent, real, and positive.

We derive equations of motion for the operators $\psi_{k}({\bf
x})$ and $c$. First we study laser cooling of the phase $\Phi$ in
a semiclassical model before returning to the full quantum model
below. In the semiclassical model we assume $N \gg 1$ and replace
the operators $\psi_{k}({\bf x})$ and $c$ by the c-numbers
$\Psi_{k}({\bf x})$ and $C$, respectively. We define $C_1=C$, 
$C_2=C^*$, denote the Kronecker delta by $\delta_{k1}$ and find
\begin{subequations}
\label{mo}
\begin{align}
\dot \Psi_{k} =&-i\left(-\frac{\nabla^2}{2 m}+V +\sum_{l}u_{kl}
\left|\Psi_{l}\right|^2 \right)
\Psi_{k}\nonumber \\
& \quad -i\left(g C_{k} +\frac{\Omega}{2}\right)\Psi_{l\neq k}
-i\Delta
\delta_{k1} \Psi_{k}, \label{mo1} \\
\dot C =& -i \int \! d{\mathbf x} \; g \Psi_2^* \Psi_1 - \left(i
\nu + \frac{\kappa}{2}\right) C. \label{mo3}
\end{align}
\end{subequations}
In terms of this semiclassical framework the initial independence
of the two condensates is modelled by a random initial phase,
i.e.\ $\rho_{kl}({\mathbf x} ,{\mathbf x}') \approx \langle
\Psi_{l}^*({\mathbf x}') \Psi_{k}({\mathbf x})
\rangle_{\rm cl}$. Here $\langle \dots \rangle_{\rm cl}$ is the
stochastic average over relative phases $\Phi$ which are initially
equally distributed between $[0,2\pi]$. Apart from the relative
phase the initial conditions for solving Eqs.~(\ref{mo}) are
assumed to be identical.

For analytical calculations our numerical studies (see below)
suggest the two mode ansatz (with $\varphi({\mathbf x})$ a fixed
and normalized wave function)
$ \Psi_{k}({\mathbf x}) = \sqrt{N_{k}}\varphi({\mathbf x}){\rm exp}
{(i\Phi_{k})}$, where $N_{k}$ is the expectation value of the number 
of particles in state $|k\rangle$, and $\Phi_{k}$ is an ${\mathbf x}$
independent total phase. Within this semiclassical two mode model
we find:
\begin{subequations}
\begin{align}
\dot{\Phi} =& \mu_2-\mu_1-\Delta+
\left(\sqrt{\frac{N_1}{N_2}}-\sqrt{\frac{N_2}{N_1}}
\right) \; {\rm Re}\{\alpha\}, \\
\dot{P}_{\Phi} =& - 2 \sqrt{N_1 N_2} \; {\rm Im} \left\{ \alpha \right\}, \\
\dot{C} =& -i\sqrt{N_1 N_2} g e^{i\Phi}-\left(i\nu
+\frac{\kappa}{2}\right)C,
\end{align}
\label{pheq}
\end{subequations}
where we have defined $\Phi=\Phi_{1}-\Phi_{2}$,
$P_{\Phi}=(N_2-N_1)/2$ and $\alpha=(gC+\Omega/2)\exp(-i\Phi)$. The
chemical potentials $\mu_k$ are defined by \begin{equation} \mu_{k}= \int \!
d{\mathbf x} \; \varphi \left(-\frac{\nabla^2}{2 m}+
V+\sum_{l}u_{kl} N_l \left|\varphi \right|^2 \right) \varphi. \end{equation}

To make contact with the resistively shunted junction model
\cite{JJ}, often used in describing JJs we set $u_{11}=u_{22}$,
and assume $N \gg 2|P_{\Phi}|$ and $NU \gg \{\Omega,gC\}$, which 
can be enforced by adjusting the laser parameters correspondingly.
Here $U=(u_{11}-u_{12})\int d{\mathbf x}|\varphi({\mathbf x})|^{4}$. We
further simplify Eqs.~(\ref{pheq}) by adiabatically eliminating
the cavity mode $C$ to second order, valid when $\kappa \gg \{g
\sqrt{N}, UN\}$, and derive an intuitively appealing damping
equation for a fictitious particle moving along a coordinate
$\Phi$ (with $M=1/2U$),
\begin{equation}
M \ddot \Phi + a_r \dot\Phi + \frac{\Omega N}{2} \sin(\Phi)=F_d.
\label{Phaseev}
\end{equation}
The particle moves in a potential $V(\Phi)=\Omega N \cos(\Phi)/2 +
F_d \Phi$ where $F_d=g^2 N^2 \kappa/(\kappa^2+4\nu^2)$ is a
constant force. This potential has minima at
$\Phi=(2n+1)\pi-\arcsin(\Omega_c/\Omega)$ with integer $n$ for
$\Omega > \Omega_c= 2 F_d /N$. Close to a minimum $V(\Phi)$ is
harmonic with frequency $\bar \omega^2=UN
\sqrt{\Omega^2-\Omega_c^2}$. The motion of the particle in
$V(\Phi)$ is damped with a friction coefficient $a_r=\nu g^2 N^2
\kappa/ 2 \left(\kappa^2/4 + \nu^2 \right)^2$. From
Eq.~(\ref{Phaseev}) we easily find the damping time scale of the
relative phase $\Phi$, which is given by  $\tau=2 M/a_r$. Note
that cooling of the phase towards a minimum of the potential
occurs only for $a_r>0$ which requires a cavity detuning of
$\nu>0$ and that both $F_d$ and $a_r$ are induced by the cavity
damping rate $\kappa$.

If $NU \leq \{\Omega,gC\}$ Eq.~(\ref{Phaseev}) no longer
holds. However, from Eqs.~(\ref{pheq}) we can still find an
approximate expression for $\tau$ which has a minimum at
$\Omega=\Omega_{m}=[(4 \nu^2 -\kappa^2+ 2\sqrt{\kappa^4+4\nu^2
\kappa^2 +16 \nu^4})/12]^{1/2}$. This case corresponds to the
Rabi-oscillation limit \cite{JJ}.

\begin{figure}
\begin{center}
\includegraphics[width=7.7cm]{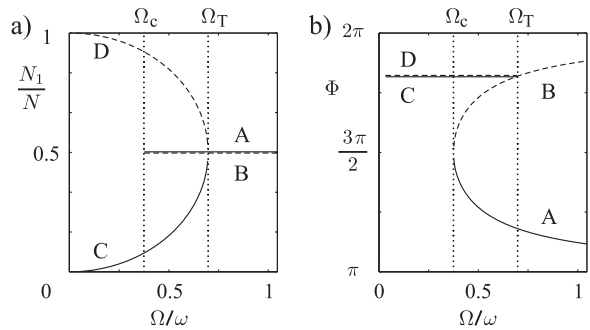}
\end{center}
\caption{Mean particle number $N_1$ in state $|1\rangle$ [a)] and
phase $\Phi$ [b)] of the condensate in the stationary state
against Rabi-frequency $\Omega$. The solid (dashed) curves
correspond to stable (unstable) stationary solutions. The dotted
vertical lines give the critical Rabi frequencies $\Omega_{c}$ and
$\Omega_T$ as defined in the text. The parameters are $UN=0.213
\omega$, $\Delta=0$, $\sqrt{N} g=3.87 \omega$, $\kappa=40 \omega$
and $\nu=20 \omega$, corresponding to condensates of total
$N=6000$ Rb atoms with $m=1.44 \cdot 10^{-25}$kg and scattering
lengths of $a_1=a_2=5.45nm$, $a_{12}=5.24nm$ trapped in a harmonic
potential with frequency $\omega=200 {\rm Hz} / 2\pi$.}
\label{fig2}
\end{figure}

We now determine the stationary solutions of our semiclassical two
mode model [Eqs.~(\ref{pheq})] which are characterized by $\dot C
= 0$, $\dot P_{\Phi}=0$ and by $\dot \Phi=0$. Writing  $C=|C|
\exp(i \Phi_c)$ we find the following conditions:
\begin{subequations}
\begin{align}
2 g  \sqrt{N_1 N_2} \sin(\Phi - \Phi_c) =& \kappa |C|, \\
-g \sqrt{N_1 N_2} \cos(\Phi - \Phi_c) =& \nu |C|,\\
-\Omega \sqrt{N_1 N_2} \sin \Phi =&\kappa |C|^2,\\
\sqrt{N_1 N_2} \left(\mu_1+\Delta-\mu_2\right) =& (N_1-N_2) {\rm
Re}\{\alpha\}.
\end{align}
\label{stat1}
\end{subequations}
We set the macroscopic wave function $\varphi({\bf
x})=\varphi_0({\bf x})$ where $\varphi_0({\bf x})$ is found from
numerically solving Eqs.~(\ref{mo1}) for $N_1=N_2$ with all the
lasers turned off ($\Omega=g=\Delta=0$) for the ground state. Thus
we find $\mu_2-\mu_1= U (N_2-N_1)$. As shown in Fig.~\ref{fig2} we
obtain one stable (unstable) branch A (B) of stationary solutions
with $N_1=N_2$ for $\Omega>\Omega_{c}$. For $\Omega<\Omega_T$,
where $\Omega_T^2=\Omega_{c}^2 + [N U + N\nu g^2 / (\nu^2
+\kappa^2/4)]^2$, there exist stationary solutions with $N_1\neq
N_2$ where the solutions on branch C (D) with $N_1<N/2$
($N_1>N/2$) are stable (unstable). The solutions of branches A and
B can also be found from the Josephson model Eq.~(\ref{Phaseev})
while branches C and D are beyond the range of validity of
Eq.~(\ref{Phaseev}). The stability of the stationary
solutions is checked by a linear stability analysis of
Eqs.~(\ref{pheq}). We also numerically evolve Eqs.~(\ref{mo})
with initial using the two mode ansatz where $N_1$,
$N_2$, $\Phi$, are found from Eqs.~(\ref{stat1}) and $\varphi({\bf
x})=\varphi_0({\bf x})$. In numerically solving Eq.~(\ref{mo}) we
restrict ourselves to one spatial dimension \cite{1Dpapers}. For
solutions on branch A these initial conditions are exact
stationary solutions of Eqs.~(\ref{mo}) and we find them to be
numerically stable. The phase $\Phi$ always evolves towards its
stationary value. Therefore solutions on branch A are best suited
for uniting two condensates. On branch C with $N_1 \neq N_2$ which
is less interesting for the purpose of laser cooling of the phase
$\Phi$ the validity of the two mode ansatz depends on the
strength of the two particle interaction. For $UN < \omega$ we
find good agreement between the numerics and our analytical
results while for $UN > \omega$ instabilities in the evolution of
Eqs.~(\ref{mo}) emerge which do not appear in the linear stability
analysis of Eqs.~(\ref{pheq}).

\begin{figure}
\begin{center}
\includegraphics[width=7.7cm]{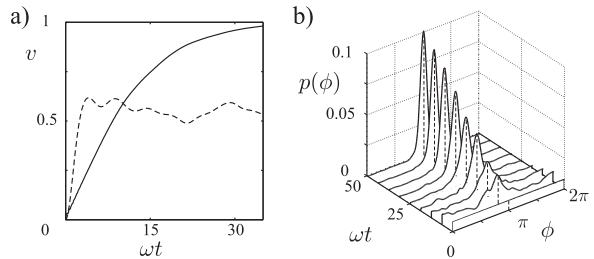}
\end{center}
\caption{a) Visibility $v$ in the Ramsey experiment against
cooling time $t$ for the parameters as in Fig.~\ref{fig2}:
$\Omega=15 \omega$ (solid line), $\Omega=0.55 \omega$ (dashed)
where two stable stationary solutions exist. b) Time evolution of
the phase distribution $p(\phi)$  from a Monte-Carlo simulation.
The dashed vertical lines indicate the semiclassical solution in
the stable stationary state. For parameters see Fig.~\ref{fig2},
$\Omega=40 \omega$ and  $N=50$ particles.} \label{fig3}
\end{figure}

A possible experimental scenario to monitor the locking of the
relative phase  $\Phi$ is a Ramsey experiment. After time $t$
cooling is stopped and a $\pi/2$ pulse with phase $\Phi_R$ is
applied to the condensates \cite{JILARams}. This pulse effectively
transforms the relative phase $\Phi$ into a difference in
occupation numbers. For a well defined phase $\Phi$, by varying
$\Phi_R$ we expect to observe maximum fringe visibility $v=1$,
whereas for an undefined, effectively random phase $v=0$. In terms
of the one particle density matrix $\rho_{kl}({\bf x},{\bf x}')$
the visibility is given by $v({\bf x})=2 |\rho_{12}({\bf x},{\bf
x})| /[\rho_{11} ({\bf x},{\bf x})+\rho_{22}({\bf x},{\bf x})]$ at
position $\bf x$. Figure~\ref{fig3}a shows the visibility
$v=v({\bf x} =0)$ in the center of a one dimensional trap found by
numerically solving the semiclassical model [Eqs.~(\ref{mo})]
against the cooling time $t$. As expected $v$ does not go to $1$
in the bistable region while it tends to $1$ for $\Omega>\Omega_T$. 
The results agree very well with the semiclassical two mode model Eqs.~(\ref{pheq}).

So far we have studied a semiclassical version of a
Josephson--cavity model, where the atoms and the cavity mode are
described by c-number fields. We will investigate now quantum
effects within a two-mode model of a JJ \cite{JJ}. This is
obtained from the Hamiltonian (\ref{H}) with the replacement
$\psi_k({\bf x})=b_k \varphi({\bf x})$ where $b_k$ are destruction
operators of a particle in electronic state $|k\rangle$, and
$\varphi({\mathbf x})$ a (fixed) spatial mode function. We define
an operator describing the imbalance in the particle numbers
$p_\phi=(b_2^\dagger b_2 - b_1^\dagger b_1)/2$, and its canonical
conjugate $-\phi$, where $\phi$ has the meaning of a relative
phase operator \cite{JJ}. Projection on the eigenstates of the
phase operator, $|\phi\rangle$, gives the quantum phase
distribution $p(\phi)={\rm tr}\{|\phi\rangle \langle \phi | \rho
\}$. We note that the eigenstates of the phase operator
$|\phi\rangle \sim \sum_m e^{-i m \phi}|N/2-m\rangle_1
|N/2+m\rangle_2$ are entangled states of particles in the first
and second internal state \cite{JJ}. For small particle numbers
and low occupation of the cavity mode, the master equation
(\ref{master}) can be solved directly by quantum Monte Carlo
techniques, and the time evolution of the phase distribution can
thus be calculated exactly. As an example, Fig.~\ref{fig3}b shows
the evolution of the phase distribution $p(\phi)$ for $N=50$
atoms, $u_{11}=u_{22}$, and initially independent condensates,
i.e. $p(\phi)=1/2\pi$, obtained from a simulation. During the
cooling process $p(\phi)$ builds up around the semiclassical
stationary solution for $\Phi$ on the time scale predicted by the
semiclassical model, and one finds excellent agreement between the
Monte Carlo simulation and the semiclassical model, even for a
small number of particles.

The quantum limits of the steady state distribution of the phase
can be discussed within a quantum phase model. For $UN \gg
\{\Omega,gc\}$ the Hamiltonian simplifies to 
\begin{equation} 
H \!= \! -U
\frac{\partial^2}{\partial \phi^2} \!- i \Delta
\frac{\partial}{\partial \phi} \!+\frac{\Omega N}{2} \cos \phi
\!+\frac{N g}{2}\left(c e^{\!-i \phi} \! + \!{\rm h.c.}
\right)\!+\nu c^\dagger c . \label{Hphase}
\end{equation}
which describes the
motion of a fictitious quantum particle in a $\cos \phi$
phase-potential coupled to a (damped) harmonic oscillator
representing the cavity mode. In this picture, cavity assisted
phase locking can be understood as ``cooling of the motion'' of
this fictitious particle in the phase potential due to coupling to
a damped harmonic oscillator. In fact, the master equation
(\ref{master}) with Hamiltonian (\ref{Hphase}) is identical to the
master equation which have been derived for laser cooling of
single atoms or molecules in a high-Q cavity moving in a trapping
potential \cite{singleatomcool}. Analytical expressions for the
width of the steady state phase distributions can be given by
expanding the Hamiltonian (\ref{Hphase}) around the semiclassical
steady state solutions to second order, and solving for the
variance of  $\phi$ and $p_\phi$ in the stationary state. We find
\cite{Omrem} 
\begin{eqnarray}
\Delta \phi^2 &=& U \frac{\kappa^4+4 \kappa^2 (\bar \omega^2 +2 \nu^2) \!-\! 16 \bar g^2 \nu
\bar \omega\!+16 \nu^2( \bar \omega^2 \!+\! 
\nu^2)}{8 \nu \bar \omega(\kappa^2 \bar \omega \!-\! 4 \bar g^2 \nu
\!+\! 4 \bar \omega \nu^2)}, \nonumber \\
\Delta p_\phi^2 &=&  \frac{\kappa^2+ 4 \nu^2 + 4 \bar \omega^2}{32
U \nu}, \end{eqnarray}
where $\bar g=N g \sqrt{U/\bar \omega}$. In the limit of cavity
width much larger than the oscillation frequency in the phase
potential, $\kappa \gg \{ \bar \omega , g\}$, the minimum
uncertainty of the phase variance is obtained for the detuning
$\nu=\kappa/2$, and is given by $\Delta \phi ^2 \approx U \kappa/2
\bar \omega^2 \gg U / \bar \omega$. On the other hand, for small
cavity damping, $\{\kappa,\bar g \} \ll \{\nu, \bar \omega \}$, we
find $\Delta \phi^2 \approx U / \bar \omega$ and $\Delta p_\phi^2
\approx \bar \omega / 4 U$ for $\nu=\bar \omega$. By analogy with laser
cooling we can identify the first limit with the {\em Doppler regime},
where a semiclassical description is valid and the final temperature is
$kT=\kappa / 2$, whereas the second case corresponds to the {\em
sideband cooling limit}, where the particle is cooled to the
vibrational ground state, i.e. we have a minimum uncertainty state
($T=0)$.

The above results were obtained for space independent couplings
$g(\mathbf x)=g$ and $\Omega(\mathbf x)=\Omega$. However, due to
imperfections absorption and emission of photons may lead to a
small momentum transfer to the condensates resulting in spatial
perturbations to the positional condensate mode $\varphi({\mathbf
x})$. A detailed analysis shows that these perturbations are
damped away with a cooling mechanism similar to the phase cooling
described above \cite{SimonCool}.

In summary, we have shown that by coupling two condensates to a
dissipative ring cavity we can ``laser cool'' the relative phase
between two initially independent condensates without loss of
particles.

We thank R. Ballagh, R. Dum, P. Horak, and H. Ritsch for
 discussions. Research supported by the Austrian and European Science Foundation, and by the TMR
networks ERB-FMRX-CT96-0087 and ERBFMRX-CT96-0002.


\begin{thebibliography}{99}
\bibitem{Omrem}
The frequency $\bar \omega^2$ is shifted from the previously
defined value by $4 g^2 U N^2 \nu /(\kappa^2 + 4 \nu^2)$.
\end{thebibliography}
\end{document}